\documentclass[prl,superscriptaddress,twocolumn,amsmath,amssymb,nofootinbib,longbibliography]{revtex4-2}

\usepackage{graphicx}
\usepackage{amsmath}
\usepackage{physics}
\usepackage{color}
\usepackage{xcolor}
\usepackage{slashed}
\usepackage{tensor}
\usepackage[utf8]{inputenc}
\usepackage[normalem]{ulem}

\newcommand{\ba}{\begin{eqnarray}}
\newcommand{\ea}{\end{eqnarray}}

\newcommand{\be}{\begin{equation}}             
\newcommand{\ee}{\end{equation}}               
\newcommand{\notprop}{\propto\kern-1\@ptsize pt \diagup}

\begin{document}

\title{Multicritical Phase Transitions in Multiply Rotating Black Holes}

\author{Jerry Wu}
\email{yq4wu@uwaterloo.ca} 
	\affiliation{Department of Physics and Astronomy, University of Waterloo,
		Waterloo, Ontario, Canada, N2L 3G1}
	
\author{Robert B. Mann}
	\email{rbmann@uwaterloo.ca}
		\affiliation{Department of Physics and Astronomy, University of Waterloo,
		Waterloo, Ontario, Canada, N2L 3G1}	
	\affiliation{Perimeter Institute, 31 Caroline Street North, Waterloo, ON, N2L 2Y5, Canada}

\date{\today}
	
	\begin{abstract}
		We show that multi-critical points in which more than three phases coalesce are present in multiply rotating Kerr-AdS black holes in $d$-dimensions. We explicitly present a quadruple point for a triply rotating black hole in $d=8$ and a quintuple point for a quadruply rotating black hole in $d=10$. 
  The maximal number of distinct phases $n$ is one larger than the maximal number of independent rotations, and   we   outline a method for obtaining the associated  
  $n$-tuple point. 
 Situations also exist where more than three phases  merge at sub-maximal multi-critical points.  Our results show that multi-critical points in black hole thermodynamics  are more common than previously thought, with systems potentially supporting many phases as long as a sufficient number of thermodynamic variables are present.  
	\end{abstract}

\maketitle

Black hole thermodynamics provides crucial guidance along the path toward a quantum theory of gravity.
Asymptotically anti de Sitter (AdS) black holes have been of particular importance to this end ever since the discover of 
a phase transition between thermal radiation and a large AdS black hole
(known as the Hawking-Page (HP) transition ~\cite{Hawking:1982dh}), which corresponds to the confinement/deconfinement of a dual quark gluon plasma \cite{Witten:1998zw} in the context of the AdS/CFT correspondence.

Once it was understood that a cosmological constant $\Lambda$  
can be understood  
as a thermodynamic variable (associated, for example, with a $(d-1)$-form gauge field \cite{Creighton:1995au}) corresponding to pressure~\cite{Caldarelli:1999xj,Kastor:2009wy,Cvetic:2010jb}, black holes were seen to exhibit 
a broad range of phase behaviour.  Charged black holes can undergo Van der Waals transitions \cite{Kubiznak:2012wp},  the Hawking-Page transition can be understood as a solid-liquid transition \cite{Kubiznak:2014zwa}, 
reentrant transitions take place in rotating black holes \cite{Altamirano:2013ane}, scalar couplings admit superfluid transitions \cite{Hennigar:2016xwd}, Lovelock black holes can have  polymer-type phase transitions \cite{Dolan:2014vba}, and accelerating black holes \cite{Anabalon:2018qfv} have snapping transitions   in which Van der Waals behaviour suddenly disappears
\cite{Abbasvandi:2018vsh}.   The resemblance of these phenomena to chemical phase transitions has prompted a
molecular interpretation of the underlying constituent degrees of freedom \cite{Wei:2019uqg}, and the subject has come to be known as 
{\em Black Hole Chemistry} \cite{Kubiznak:2016qmn}.

The existence of triple points, in which three black hole phases coalesce at a single pressure and temperature (analogous to ice/water/steam), were discovered some time ago in doubly rotating black holes \cite{Altamirano:2013uqa}, and subsequently in Lovelock gravity \cite{Wei:2014hba,Frassino:2014pha}; more recently  a proposal for the microstructure of black holes at  such points was proposed  \cite{Wei:2021krr}.
However multi-critical points of the type seen in colloidal polymers and other heterogeneous systems
\cite{Akahane2016,Garcia2017,Sun:2021gpr}, in which more than three phases merge,  did not seem to be present in black hole physics.  Recently 
such multi-critical points were  discovered for charged AdS black holes in non-linear electrodynamics \cite{Tavakoli:2022kmo}. 

Here we show that the family of multiply-rotating Kerr-AdS black holes \cite{Gibbons:2004js,Gibbons:2004uw}  also exhibits  multi-critical behaviour.  
The different angular momenta  introduce additional thermodynamic conjugate pairs to the system, allowing for more phases 
 than the small/intermediate/large  ones  seen for doubly rotating black holes  \cite{Altamirano:2013uqa}.  We find 
 multiple phases separated by first order phase transitions  for sufficiently high pressure and appropriate angular momenta. As the pressure is lowered, these phases   merge at a single pressure and temperature; for pressures below this multi-critical point  only the largest and smallest black hole phases remain, separated by a first order phase transition.  We explicitly show the existence of
 a quadruple point for a triply-rotating black hole and a quintuple point for a black hole with 4 angular momenta.  In general we find that 
 a black hole with $(n+1)$ distinct angular momenta can have an $n$-tuple point, as well as multi-critical points of lower order.  Our results suggest that black hole  multi-critical behaviour is common, requiring neither the introduction of unusual matter sources nor a theory of gravity different from general relativity.    
  
Setting $\hbar=c=G=1$, the general metric for mutiply rotating Kerr-AdS black holes is \cite{Gibbons:2004js,Gibbons:2004uw} 
\ba \label{metric}
ds^2&=&-W\Bigl(1+\frac{r^2}{l^2}\Bigr)d\tau ^2+\frac{2m}{U} \Bigl(W d\tau -\sum_{i=1}^{N} \frac{a_i \mu_i ^2 d\varphi _i}{\Xi _i}\Bigr)^2\nonumber\\
&+&\sum_{i=1}^{N} \frac{r^2+a_i^2}{\Xi _i} \mu_i ^2 d\varphi _i^2+\frac{U dr^2}{F-2m}+\sum_{i=1}^{N+\epsilon}\frac{r^2+a_i ^2}{\Xi _i} d\mu _i ^2 \nonumber\\
&-&\frac{l^{-2}}{W (1+r^2/l^{2})}\Bigl(\sum_{i=1}^{N+\epsilon}\frac{r^2+a_i ^2}{\Xi _i} \mu_i d\mu_i\Bigr)^2\,.
\ea
in $d$ spacetime dimensions, with metric functions 
\ba\label{metrcifunctions}
W&=&\sum_{i=1}^{N+\epsilon}\frac{\mu _i^2}{\Xi _i}\,,\quad U=r^\epsilon \sum_{i=1}^{N+\epsilon} \frac{\mu _i^2}{r^2+a_i^2} \prod _j ^N (r^2+a_j^2)\,,\nonumber\\
F&=&r^ {\epsilon -2} \Bigl(1+\frac{r^2}{l^2}\Bigr) \prod_{i=1}^N (r^2+a_i^2)\,,\quad \Xi_i=1-\frac{a_i^2}{l^2} 
\ea
and where   $n=\frac{1}{2}(d-1-\epsilon)$, is  the maximal number of independent rotations, with $\epsilon=0/1$ for odd/even spacetime dimensions.
The coordinates $\mu_i$ obey the constraint $\sum_{i=1}^n \mu^2_i = 1$. 
  $l$ is the AdS radius, $m$ is  the mass parameter, and $a_i$ are the rotation parameters. The horizon radius $r_+$ can be determined by finding the largest root of the equation $F-2m=0$. For constant $J_i$, the $a_i$ parameters are functions of $r_+$ and  the thermodynamic pressure $P$, where
\be
P=\frac{-\Lambda}{8\pi}, \qquad \Lambda=-\frac{(d-1)(d-2)}{2 l^2}
\ee
in Planckian units $\ell_P^2 = \frac{G\hbar}{c^3}$ \cite{Kubiznak:2016qmn}. 
The mass $M$, angular momenta $J_i$, and the thermodynamically conjugate angular velocities $\Omega_i$ are \cite{Gibbons:2004ai}
\ba \label{TD}
M&=&\frac{m \Sigma _{d-2}}{4\pi (\prod_j \Xi_j)}(\sum_{i=1}^{N}{\frac{1}{\Xi_i}-\frac{1-\epsilon }{2}})\,,\nonumber\\
J_i&=&\frac{a_i m \Sigma _{d-2}}{4\pi \Xi_i (\prod_j \Xi_j)}\,,\quad \Omega_i=\frac{a_i (1+\frac{r_+^2}{l^2})}{r_+^2+a_i^2}\,,
\ea
where $\Sigma _{d-2} = \frac{2\pi^{\frac{d-1}{2}}}{\Gamma(\frac{d-1}{2})}$. The temperature $T=\frac{\kappa}{2 \pi}$ and entropy $S$ are determined in terms of   $r_+$
\ba\label{TS}
T&=&\frac{1}{2\pi }\Bigr[r_+\Bigl(\frac{r_+^2}{l^2}+1\Bigr)
\sum_{i=1}^{N} \frac{1}{a_i^2+r_+^2}-\frac{1}{r_+}
\Bigl(\frac{1}{2}-\frac{r_+^2}{2l^2}\Bigr)^{\!\epsilon}\,\Bigr]\,,\nonumber\\
S&=&\frac{\Sigma _{d-2}}{4 r_+^{1-\epsilon}}\prod_{i=1}^N 
\frac{a_i^2+r_+^2}{\Xi_i}\,.
\ea
The thermodynamically stable state of the system is given by the global minimum of   the Gibbs free energy $G=M-TS$ \cite{Kubiznak:2016qmn}.
 
\begin{figure}
	\includegraphics[width=0.48\textwidth]{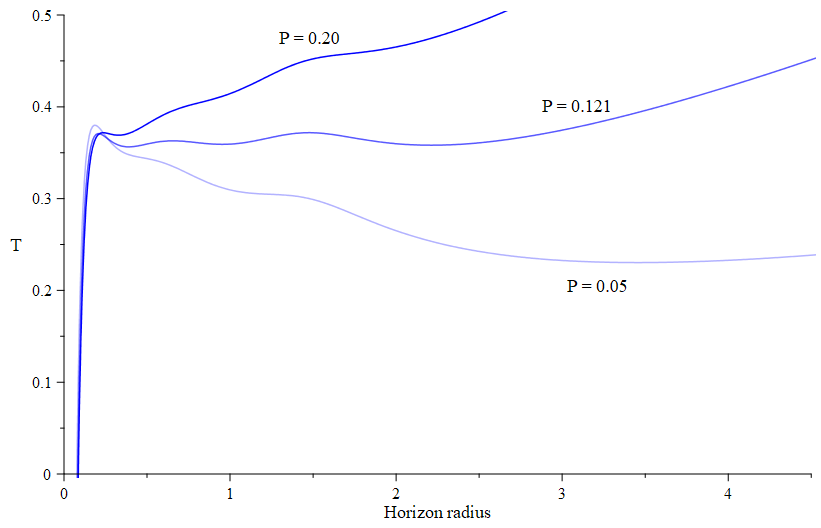}
	\includegraphics[width=0.48\textwidth]{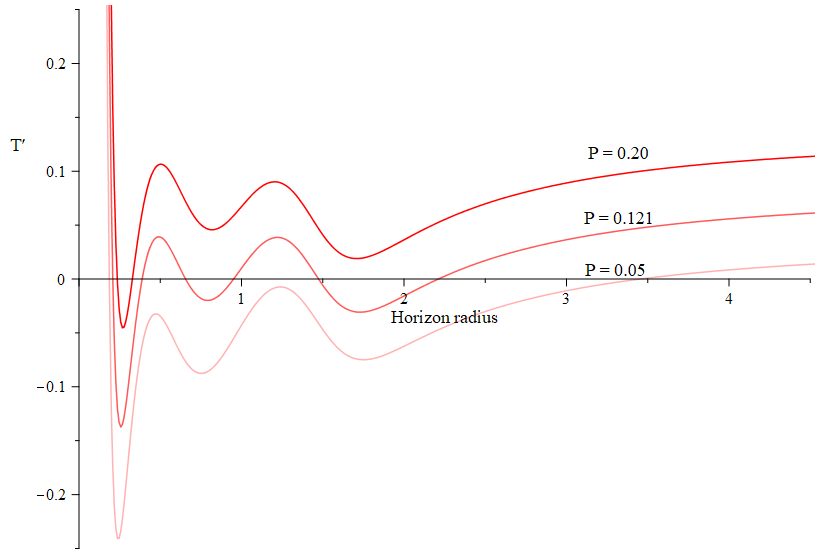}

	\caption{\textbf{$T$ and $T^\prime$ for different values of $P$.} $d=8$, $J_1 = 7.967$, $J_2 = 1.24$, $J_3 = 0.12798$. 
	\textit{Top.} $T$ at three different pressures, attaining local extrema at the roots of $T^\prime$. A maximum of three pairs of maxima and minima is displayed for $P=0.121$, indicating three swallowtails. \textit{Bottom.} The derivative $T^\prime$ at these three pressures, showing different numbers of roots. 
	}
	\label{fig:3rotation_TTprime} 
\end{figure}

The  first law is  \cite{Gibbons:2004ai,Altamirano:2014tva}
\be \label{firstlaw}
dM=TdS + \sum_{i=1}^{N} \Omega_i dJ_i + V dP
\ee
with the thermodynamic volume
\be
V = \frac{r_+ A}{d-1} + \frac{8\pi}{(d-1)(d-2)}\sum_{i=1}^n a_i J_i\; .
\ee
Taking the variation of $G$ and substituting \eqref{firstlaw} yields
\begin{align}
      dG&=\sum_{i=1}^{N} \Omega_i dJ_i - SdT -V dP
\end{align}
and so  $dG=-SdT$ for constant $P$ and $J_i$, in which case the extrema of $G(r_+)$ and $T(r_+)$ occur at the same $r_+$ values. Consequently $T$ alone determines the existence and distribution of swallowtails in the $G$-$T$ plot, with the cusps of the swallowtails corresponding to the zeros of $T^\prime = \frac{\partial T}{\partial r_+}$.
It is possible to choose constant $J_i$ so that $T^\prime$ has a new set of local maxima and minima for each new rotation. Adjusting the pressure changes the locations of these extrema;  the maximum number of coexistent states is attained when $T^\prime(r_+)$ has a root between every local extremum. Unlike the doubly rotating case \cite{Altamirano:2013uqa} where the system only depends on the ratio between the two angular momenta, in general the $a_i$ rotational parameters are not invariant under constant scaling of $J_i$, and so we will directly fix the $J_i$.

We begin by illustrating the existence of a quadruple point for a triply rotating black hole in $d=8$, the minimal value needed to   support 4 distinct phases. 
\begin{figure}
	\includegraphics[width=0.48\textwidth]{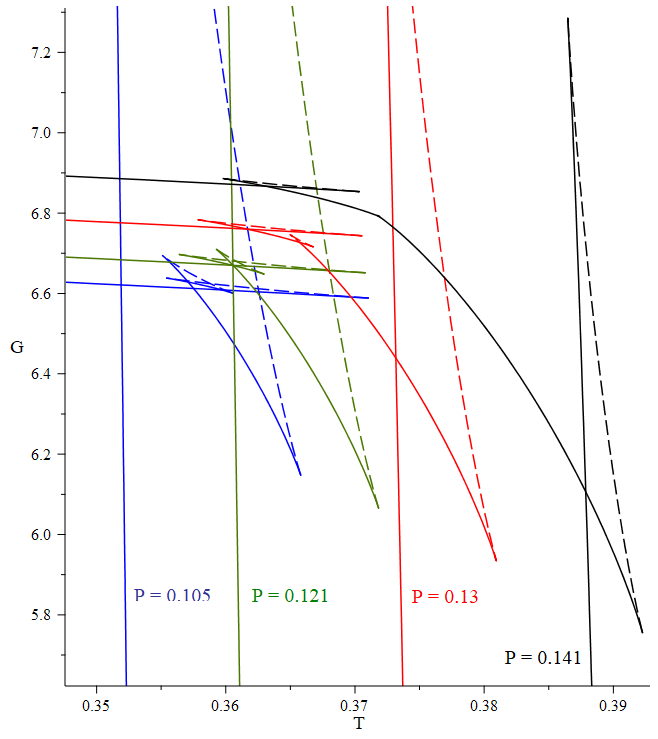}

	\caption{\textbf{G-T plot with three rotations.} $d=8$, $J_1 = 7.967$, $J_2 = 1.24$, $J_3 = 0.12798$. At $P=0.141$ (black curve), only two swallowtails exist	
indicating two stable phase transitions between three distinct phases. As the pressure is lowered a third swallowtail appears (red curve) signify four distinct phases. At   $P_q=0.121$, these three swallowtails merge at a quadruple point (green curve). For pressures lower than $P_q$, only one stable first order phase transition is seen (blue curve). Dashed lines indicate negative specific heat.}
	\label{fig:3rotation_GT} 
\end{figure}
The temperature is 
\begin{align}
T&=\frac{1}{2\pi } \biggr[r_+ \left(1 + \frac{8 r_+^2 \pi P}{21}\right) \sum_{i=1}^3 \frac{1}{a_i^2 +r_+^2} \nonumber \\
&- \frac{1}{r_+} \left(\frac{1}{2} - \frac{4 r_+^2 \pi P}{21}\right)\biggl]
\end{align}
where  the $a_i$ are calculated numerically as functions of $r_+$ using \eqref{TD}, as an analytic solution is not possible.  All of our results
satisfy the constraint $|a_i| < l$, so that the metric \eqref{metric} remains well-defined.

\begin{figure}
	\includegraphics[width=0.48\textwidth]{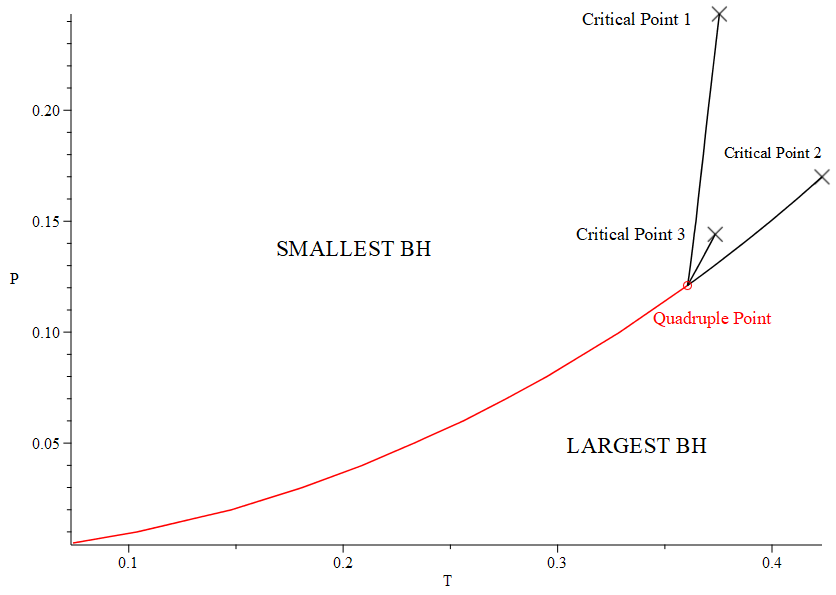}

	\caption{\textbf{P-T phase diagram for quadruple point with 3 rotations.} $d=8$, $J_1 = 7.967$, $J_2 = 1.24$, $J_3 = 0.12798$. For low pressures only one stable phase transition between the smallest and largest black holes   exists (red curve). At $P=P_q$, four phases (smallest, small, large, largest) coexist at $T_q\approx0.3606$. For $P_q < P < P_{c_3}$, three stable first order phase transitions are observed between four phases. All three coexistence curves terminate at their respective critical points.
	}
	\label{fig:3rotation_PT} 
\end{figure}

For large $P$, no phase transitions are present. As $P$  decreases, $T(r_+)$ develops extrema (shown in  figure~\ref{fig:3rotation_TTprime}) and 
 new thermodynamic phenomena arise illustrated in figure~\ref{fig:3rotation_GT}. 
 At the critical pressure $P_{c_1} \approx 0.24335$, a first order phase transition between the smallest black hole and a larger black hole appears
 and the $G$-$T$ plot has a single swallowtail. 
 Further lowering the pressure, two more critical pressures emerge, one at $P_{c_2}\approx 0.169948$ (where two swallowtails appear in
 figure~\ref{fig:3rotation_GT}) and then at  $P_{c_3} \approx 0.144097$ (where a third swallowtail emerges as shown in the red curve in 
 figure~\ref{fig:3rotation_GT}).  Below this latter pressure 
 four phases appear separated by three first order phase transitions between $P_{c_3}$ and $P_q = 0.121$. At $P=P_q$, the three swallowtails merge at a single  quadruple critical point where all four phases coexist. For $P<P_q$ there is only one first order phase transition between the largest and the smallest black hole; all other swallowtails  are in a thermodynamically unstable region and eventually disappear for smaller $P$.  
The coexistence  curves for figure~\ref{fig:3rotation_GT} are given in
figure~\ref{fig:3rotation_PT}.

While there is a finite range of values of $J_i$ that admit quadruple points, in general these will not be present 
for most choices of $J_i$. Instead for sufficiently low pressures the  four distinct phases will each be separated by first-order phase transitions,
and the four branches of the coexistence plot merge in two places at  two distinct triple points.

\begin{figure}
	\includegraphics[width=0.48\textwidth]{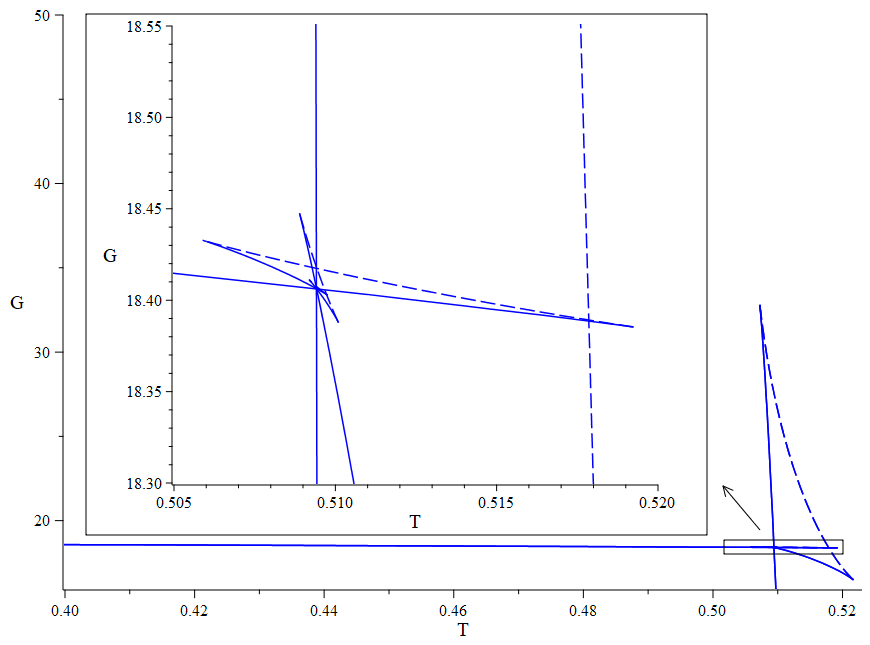}

	\caption{\textbf{G-T plot of quintuple point with 4 rotations.} $d=10$, $J_1 = 24.48$, $J_2 = 4.33$, $J_3 = 1.2$, $J_4 = 0.1435$. Four swallowtails in the Gibbs free energy merge at one point for $P=0.231$. Dashed lines indicate negative specific heat.
	}
	\label{fig:4rotation_GT} 
\end{figure}

Similarly, with appropriate choices of $J_i$, a quadruply rotating black hole in $d=10$ can support 5 distinct phases at a fixed pressure. With the added $a_4$ parameter in $T(r_+,P,J_1,J_2,J_3,J_4)$, $T^\prime(r_+)$ can admit another pair of extrema, which can then be adjusted to line up with the other three pairs seen for $d=8$. $P$ is chosen in the same way so that a root appears between each extremum of $T^\prime$, yielding five distinct black hole phases and the maximal number  of 4 swallowtails, indicating 4 first order phase transitions, which appear in the $G$-$T$ diagram for a range of fixed $P$. These merge at sufficiently small $P$, shown in
figure~\ref{fig:4rotation_GT}. The $P$-$T$ behaviour  is analogous to that of the quadruple point: one stable phase transition is present for small pressures, above which the coexistence curve splits into four branches at the quintuple point, each branch terminating at a distinct critical point as $P$ is increased. 

\begin{figure}
	\includegraphics[width=0.48\textwidth]{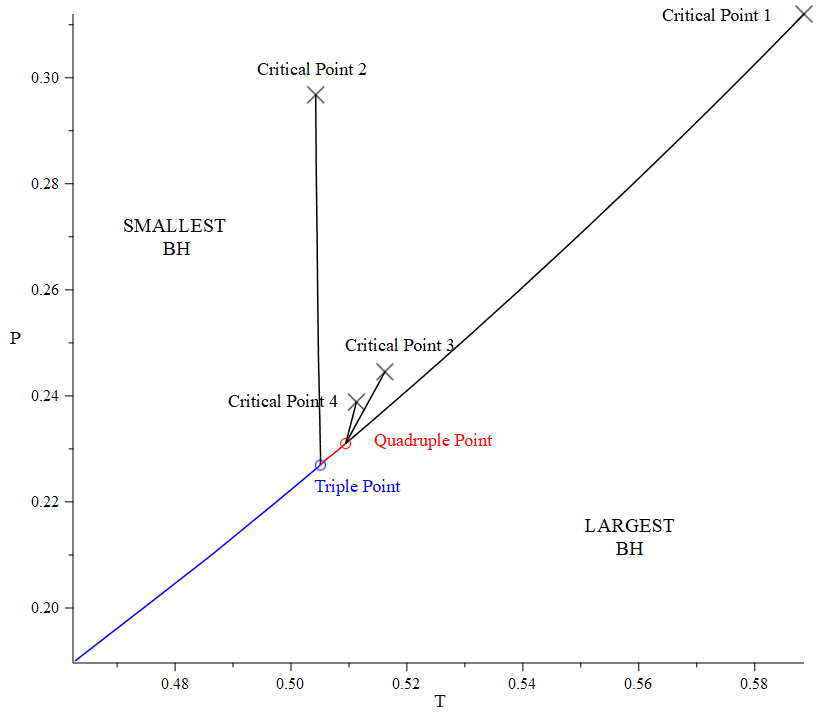}

	\caption{\textbf{P-T phase diagram with 4 rotations.} $d=10$, $J_1 = 24.48$, $J_2 = 4.331$, $J_3 = 1.1973$, $J_4 = 0.155$. For $P<P_{tr} \approx 0.22695$, the only stable phase transition observed is one of first order between the largest and smallest black hole phases (blue).  At $P=P_{tr}$, a new phase emerges at a triple point, above which three phases are present.   Two additional coexistence curves appear on the right branch at a quadruple point ($P=P_q=0.231$).
    Five distinct black hole phases exist for $P\in (P_{q}, P_{c_4} \approx 0.23883)$. All coexistence curves terminate at critical points as the pressure is increased. 
	}
	\label{fig:4rotations_tri_quad_PT} 
\end{figure}

More generally the 5 phases can exist at   fixed pressure without the presence of a quintuple point. In this case, the $P$-$T$ phase diagram can exhibit three triple points or a quadruple point and a triple point for particular choices of $J_i$. Figure~\ref{fig:4rotations_tri_quad_PT} shows a system with one quadruple point and a triple point at a lower pressure.

Multi-critical behaviour will be present for any multiply rotating AdS black hole for an appropriate
choice of angular momenta.
In general, even spacetime dimensions support up to $\frac{d}{2} - 1$ independent rotations, yielding $n = \frac{d}{2}$ potential distinct phases. The necessary conditions to obtain the maximum amount of phases is unclear, but we have found numerically that the magnitudes of $J_i$ must be sufficiently spaced out. Assuming all phases are separated by first order phase transitions, the coexistence curve has at most $n-1$ branches in regions where all $n$ phases remain, and only one branch as $P\to 0$. All $n-1$ branches of the curve eventually merge to a single branch as pressure is decreased, which can be in the form of an $n$-tuple point, or many other lower order multi-critical points at different pressures. Scaling the $J_i$ by a constant factor does not change the phase behaviour, but rather shifts the locations of critical points in the $P$-$T$ phase diagram. With sufficiently large scaling, multi-critical points can be pushed arbitrarily close to $P=T=0$. 

Multi-critical points in multiply rotating black holes are unlike those found in the context of non-linear electrodynamics \cite{Tavakoli:2022kmo} with  regards to the Gibbs phase rule, which relates the degrees of freedom $\textsf{F}$ in a simple thermodynamic system to the number of coexistence phases  $\textsf{P}$ and the number of chemical constituents. The generalized Gibbs phase rule \cite{Sun:2021gpr}:
\be
\textsf{F}=\textsf{W}-\textsf{P}+1.
\ee
 replaces the notion of chemical constituents with the number of thermodynamic conjugate pairs $\textsf{W}$, which is directly applicable in the context of black hole thermodynamics. The $n$-tuple points in non-linear electrodynamics  were discovered to have at minimum $n$ degrees of freedom and required 2 additional conjugate pairs for each new phase \cite{Tavakoli:2022kmo}, whereas in the Kerr-AdS case the $n$-tuple points  always have a lower bound of $\textsf{F}=2$, and only one added rotation is needed for a new phase. This disparity is likely due to the fact that $T$ 
 depends nonlinearly on the angular momenta $J_i$, in contrast to its linear dependence on the coupling constants in non-linear electrodynamics. 
 
 The presence of multi-critical behaviour of vacuum black holes in $d$-dimensional Einstein gravity raises a number of interesting questions.  It would be of great interest to know the necessary and sufficient conditions for multi-criticality to occur, and to find the conditions (if any) under which 
 it is possible for multiple phases to coalesce as pressure is increased.  It would likewise be interesting to understand the implications of multi-criticality for the microstructure of black holes.

\section*{Acknowledgements}
\label{sc:acknowledgements}

This work  supported in part by the Natural Sciences and Engineering Research Council of Canada (NSERC).  Perimeter Institute and the University of Waterloo are situated on the Haldimand Tract, land that was promised to the Haudenosaunee of the Six Nations of the Grand River, and is within the territory of the Neutral, Anishnawbe, and Haudenosaunee peoples.

\bibliographystyle{JHEP}

\providecommand{\href}[2]{#2}\begingroup\raggedright\endgroup

\end{document}